\documentstyle[aps,pre,twocolumn,epsfig]{revtex}
\begin{document}
\twocolumn[\hsize\textwidth\columnwidth\hsize\csname@twocolumnfalse\endcsname
\title{Polymer transport by laminar flows}
\author{M.~De Lucia$^{1}$, A.~Mazzino$^{2,3}$ and A.~Vulpiani$^{1}$ \\
\small{$^{1}$ Dipartimento di Fisica, and Istituto Nazionale di Fisica 
della Materia (UdR and  SMC), 
Universit\`a di Roma  ``La Sapienza'', Piazzale Aldo Moro 2, 
I-00185 Roma, Italy.}\\
\small{$^{2}$ CNR-ISIAtA, Str. Prov. Lecce - Monteroni Km 1.200, 
Polo Scientifico dell'Universit\`a, 73100 Lecce - Italy.}\\
\small{$^{3}$ INFM--Dipartimento di Fisica, Universit\`a di Genova,
Via Dodecaneso, 33, I-16146 Genova, Italy.}}
\draft
\date{\today}
\maketitle
\begin{abstract}
Polymer transport is investigated for two paradigmatic laminar
flows having open and closed streamlines, respectively. 
For both types of flows we find transport depletion  owing to the action
of the polymers elastic degree of freedom. For flows with closed
streamlines the leading mechanism for the observed transport reduction
is the (dynamical) formation of barriers. For flows with open 
streamlines the reduction of transport is induced by the renormalization
of the bare diffusion coefficient. Results have been obtained
by means of Lagrangian simulations.
\end{abstract}]

The transport of particles advected by a given velocity field is a subject
attracting increasing attention of physicists. 
Such issue has obvious applications in geophysics
(e.g. pollutants dispersion) as well as in industrial processes 
(e.g. transport of a solute in porous media) \cite{M83}. In addition, 
the investigation of the dynamics of
particles leads, in a natural way, to highly non trivial behaviors as the
anomalous dispersion and the Lagrangian chaos \cite{Z83,BJPV98}.

In this paper we investigate the transport 
of passive polymers in given velocity
fields. The interest of this problem is in the spectacular effect of turbulent
reduction by the dilute polymer addities (se so-called drag reduction 
\cite{B78}). 
Of course, the complete treatment of
the drag reduction is very complex and must involve the back-reaction of the
polymer concentration on the velocity field ruled by the 
Navier-Stokes equations \cite{WB88,O92}.

A first crude, but non trivial, step is to consider the passive limit, i.e.
the transport of polymers in a prescribed velocity field. This problem, in 
spite of its poor relation with the turbulent drag reduction, is not simple
at all \cite{C00,BFL00}. It can be seen as a complication of the, 
already difficult, issue of the passive particle transport \cite{M83}.

A widely used model for the polymer dynamics, due to Rouse \cite{R53,DE86}, 
considers  macromolecules as a series of $N$ beads linearly connected by 
harmonic springs.
The evolution law is given by the Langevin equations:
\begin{equation}
\zeta { d{\bf R}_n \over {dt}} = -
{\partial V \over {\partial {\bf R}_n}}  + {\bf u}({\bf R}_n) +\bbox{\eta}_n
\qquad n=0,\cdots,N-1 
\label{rouse}
\end{equation}
where $V=K/2 \sum_{n=1}^{N-1} ({\bf R}_n -{\bf R}_{n-1})^2$, ${\bf R}_n$
is the position of the $n$-th bead, ${\bf u}$ is the incompressible velocity
field, $\zeta$ is the friction coefficient, $K=3 k_B T/b^2$ is the
spring constant, $b$ is the average distance between the beads,
$k_B$ is the Boltzmann constant, $T$ is the temperature
and  $\bbox{\eta}_n$
represents the thermal noise mimicking 
the interaction of beads with the solvent.

The Rouse model (\ref{rouse}) can be further simplified (this is the so-called
dumbbell model) just taking $N=2$: 

\begin{equation}
\zeta  { d{\bf R}_0 \over {dt} }=  K ({\bf R}_1 - {\bf R}_0)
 + {\bf u}({\bf R}_0) +{\bbox \eta}_0
\label{r1} 
\end{equation}
\begin{equation}
\zeta  { d{\bf R}_1 \over {dt} }=  K ({\bf R}_0 -  {\bf R}_1)
 + {\bf u}({\bf R}_1) +{\bbox \eta}_1  . 
\label{r2} 
\end{equation}

We assume
${\bbox \eta}_i=\sqrt{2 D_0} \tilde{\bbox{\eta}}_i$, with
$\tilde{\bbox{\eta}}_i$
normalized, zero-mean, independent white noise processes, 
and we set, without loss of generality, $\zeta=1$. 
 For a given velocity field, $D_0$ and $K$ are thus 
the only relevant parameters controling the polymer dynamics.\\
In the following we shall concentrate our attention 
on the transport properties
of the bead center of mass, the coordinates of which are denoted by
 ${\bf R}\equiv ({\bf R}_0+ {\bf R}_1)/2$, 
${\bf R}\equiv (X,Y)$.
Note that in the limit of very large $K$ the evolution equation for
${\bf R}$ reduces to
the usual equation for a fluid particle but with $D_0/2$
instead of $D_0$:
\begin{equation}
{ d{\bf R} \over {dt} }= {\bf u}({\bf R}) + \sqrt{D_0}\tilde{\bbox{\eta}} .
\label{center}
\end{equation}
As in Eqs.~(\ref{r1}) and (\ref{r2}), 
$\tilde{\bbox{\eta}}$ is a normalized, zero-mean, white noise process. \\
Therefore, for the understanding of the role of the elastic degree of freedom,
results on the diffusion process of the bead center of mass
for finite values of $K$ have to be compared with those
generated by the particle diffusion limit given by Eq.~(\ref{center}).

The model (\ref{rouse}) has been investigated, e.g., in Ref.~\cite{OB94} 
for a layered random flow, and  in Ref.~\cite{WL99}
in the presence of a non-potential 
static random flow. Antithetic conclusions arose for the 
transport properties of polymers plugged  in the two above flow classes.
In Ref.~\cite{OB94}, the layered random flow has been found to cause
an enhanced transport with respect to the single particle
diffusion problem. The opposite result (i.e., the occurrence of transport 
reduction) has been singled out
in Ref.~\cite{WL99}.
 For the flows considered in Ref.~\cite{WL99} 
the basic mechanism giving rise
to transport reduction is played by the dynamically generated
barriers, which are not present in the layered random flow \cite{OB94}.\\
The physical key role
for the barrier to emerge is originated from the competition 
between flow-originated stretching, which acts on the polymers,
and the elasticity of the polymeric structure itself. Due to the
elastic degree of freedom, unlike particles, polymers have the 
possibility to select particular regions of the flow. 
Physically speaking, polymers prefer those regions where they can 
reduce their own elastic  energy. This happens, e.g., in those
regions where the velocity  strain is large and negative.  

The issue related to the possible existence of preferred regions 
in more realistic flows raised in Ref.~\cite{WL99}. One of the main aim here
is to show, by means of numerical simulations, some examples of flows
for which transport of polymers is less effective than
particles transport. The choice for the flows falls on two paradigmatic
laminar velocity fields, the diffusive properties of which have been analyzed
in great detail in the past for what concerns particle dispersion \cite{MK99}.
To be specific, we shall focus on shear flows (where the streamlines 
are open) and cellular flows both stationary and time-dependent
(where the streamlines  are closed). As we shall see, although 
for different reasons,
such classes of  flow show transport depletion in the presence of polymers.

Let us start from the convective flows.
We investigate here polymers diffusion in a simple
model mimicking the Rayleigh--B\'enard convection \cite{SG88}.
Two-dimensional convection with rigid boundary conditions is described here by
the following stream function:
\begin{equation}
\psi (x,y,t) = \psi_0 \sin (x + B\sin\omega t ) \sin y \;\;\; ,
\label{gollub}
\end{equation}
where the periodicity of the cell is $2\pi$, and the
even oscillatory instability
\cite{CB74} is accounted for by the term $B\sin \omega t$, representing
the lateral oscillation of the rolls.
Velocity is obtained from (\ref{gollub}) by the usual relations
${\bf u}=(-\partial\psi/\partial y,\partial\psi/\partial x)$.
The capability of the simple flow (\ref{gollub}) to capture the essential
features of the convection problem is discussed in Ref.~\cite{SG88}.

The second flow we have considered is the two-dimensional Kolmogorov shear 
flow defined as:
\begin{equation}
{\bf u}=\left( u(y)\;,\; 0 \right)
\label{shear}
\end{equation}
with $u(y)=U sin(y)$.

Equations (\ref{r1}) and (\ref{r2})
have been integrated with ${\bf u}$ obtained from the flows
(\ref{gollub}) and (\ref{shear}) 
using a second-order Runge-Kutta scheme.
In what  follows, averages are extended
over different realizations and are performed by following
$\sim 10^5$ couple of particles.\\
In order to investigate  the rate of transport, a measure of
the eddy diffusivity,
$D\equiv \langle   [X(t) - X(0) ]^2\rangle/(2 t)$, for large times, $t$,
has been made, $X$ being the x-component of the bead center of mass.

Let us start the discussion of our results from the cellular flows.
The behavior of the diffusion coefficient, $D$, vs the spring constant $K$
are reported in Fig.~1 for $\omega=0$ and three different values of 
the molecular diffusivity. Dashed lines correspond to the particle 
limit analytically obtained 
in Ref.~\cite{RBDH87} (formulae (22) and (24) with $d=\pi$ and 
$\beta=1$) in the limit of small $D_0$.\\
\begin{figure}
\vspace{-0cm}
\begin{center}
\includegraphics[scale=0.40,angle=-90]{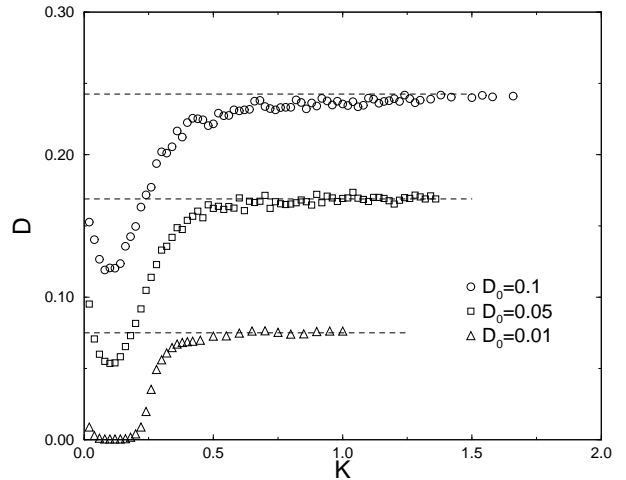}
\caption{The diffusion coefficient versus the spring constant for the 
convective model  (\ref{gollub}) with $\omega=0$ and for different values
of $D_0$. Dashed lines represent the theoretical values of the diffusion 
coefficient obtained in Ref.~\protect\cite{RBDH87}
for the particle diffusion problem.}
\end{center}
\end{figure}
In Fig.~2, the behavior of the average spring
elongation, $l\equiv \langle|{\bf R}_1-{\bf R}_0|^2\rangle^{1/2}$ 
is shown  as a function
of $K$, again for different values of $D_0$. Dashed lines correspond
to the limit of small $l$, where it is easily checked that 
 the spring elongations behaves as $(2 D_0/K)^{1/2}$.\\
\begin{figure}
\vspace{-0cm}
\begin{center}
\includegraphics[scale=0.40,angle=-90]{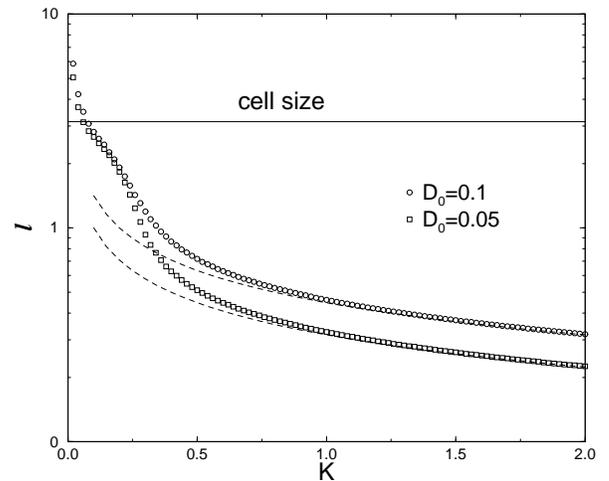}
\caption{The average spring elongation, $l$, defined as 
$l\equiv \langle|{\bf R}_1-{\bf R}_0|^2\rangle^{1/2}$, versus the spring 
constant, for different values of $D_0$. Dashed lines are relative to the 
limit of small $l$, when $l$ is found to behaves as $(2 D_0/K)^{1/2}$.}
\end{center}
\end{figure}
 From these figures,
a reduction of transport always occurs for spring 
elongations smaller than the cell size
(i.e. $l<\pi$). This is the physically interesting case.
The case $l\sim \pi$ corresponds to
 a minimum in the transport. For small values of
 $D_0$ (smaller than 
the smallest one shown in Fig.~1), there exist a narrow band
of values of $K$ corresponding to $l\sim \pi$ where 
particles remain trapped in adjacent cells. The diffusion process stops in 
this case. Note that this is a pure effect played by 
the elastic degree of freedom: 
for the single-particle diffusion in incompressible
flows, it is possible to show that the (eddy) diffusion 
coefficient is always larger
than the (bare) coefficient $D_0$\cite{MK99}.\\
Transport depletion is here caused by the aforementioned 
dynamically generated barriers. This can be easily checked in Fig.~4 
where the probability density function (pdf) for the bead
center of mass to be in the subinterval $[0,\pi] \times [\pi/4,3/4\pi]$
of the elementary cell is shown 
for different values of the spring constant $K$ (integration along $y$
has been performed).
 From this figure
it appears as polymers prefer those regions close to the
center of the cell where they can reduce their own elastic energy.
In such regions, the contribution of the molecular diffusivity
to escape from a cell to another is unimportant. On the contrary,
the regions close to the cell boundaries, where the role of molecular 
diffusivity in the (eddy) diffusion process is fundamental,
are refused by polymers. The result is   the slowing down of
the diffusing polymers. Note that the effect of barriers reduces as the
spring constant increases. For $K=0.75$, that corresponds to a spring
elongation of $l=0.40$, the probability density is almost uniform within the 
cell.\\
\begin{figure}
\vspace{-0cm}
\begin{center}
\includegraphics[scale=0.40,angle=-90]{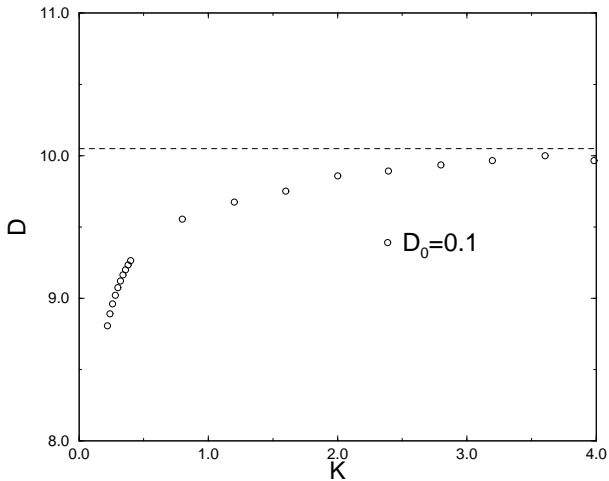}
\caption{The diffusion coefficient versus the spring constant for the 
Kolmogorov shear flow  (\protect\ref{shear}) for
$D_0=0.1$. The dashed line represents the theoretical value,
$D=(D_0/2) + U^2/[2 (D_0/2)]$, for the particle diffusion limit (see, e.g., 
Ref.~\protect\cite{BCVV95}).}
\end{center}
\end{figure}
Behaviors similar to those shown in Figs.~(1), (2) and (3) have been found
also for $\omega\neq 0$. As in Ref.~\cite{CMMV99}, we have
restricted our attention only on the case $B=0.5$.

As one can see in Fig.~3, also in the shear case the diffusion
coefficient, $D$, for finite values of $K$ 
is smaller than that in the limit case of infinite $K$
(see Eq.~(\ref{center})).
\begin{figure}
\vspace{-0cm}
\begin{center}
\includegraphics[scale=0.40,angle=-90]{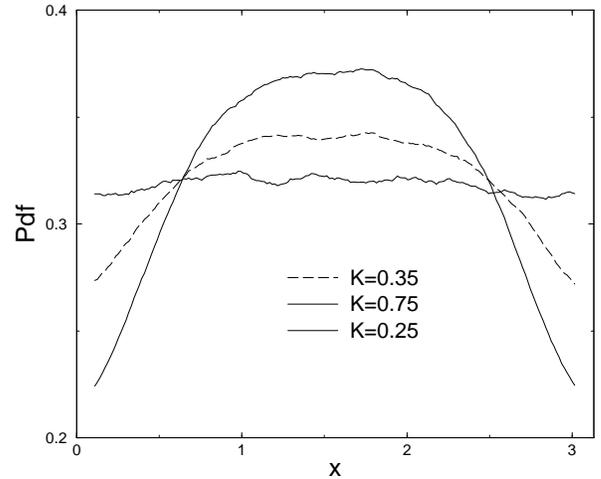}
\caption{The probability density function (pdf) for the bead
center of mass to be in the interval $[0,\pi] \times [\pi/4,3/4\pi]$
of the elementary cell. Integration along $y$ is performed.
The molecular diffusivity is $D_0=0.05$.
 For  $K=0.25$ we have $l\sim 1.37$; for $K=0.35$, $l\sim 0.75$; 
for $K=0.75$, $l\sim 0.40$.}
\end{center}
\end{figure}

Unlike what happens  in the considered cellular flows, 
this effect is not due to
the presence of barriers and can be understood by means of simple arguments.
To show that, introducing  
$\delta y\equiv Y_1-Y_0$ we immediately have from Eqs.~(\ref{r1}),
(\ref{r2}) and (\ref{shear}): 
\begin{equation}
\   { d \delta y \over {dt} }= -2K \delta y + \sqrt{4 D_0}\tilde{\eta}_1 
\end{equation}
\begin{eqnarray}
\   { d X \over {dt} }= { 1 \over 2} \left (u(Y-\delta y/2)+u(Y+\delta y/2)\right )
 + \sqrt{D_0}\tilde{\eta}_2
 \simeq\nonumber \\
u(Y)+ { {d^2u} \over {dY^2}} \frac{(\delta y)^2}{8}  + \sqrt{D_0}\tilde{\eta}_2.
\label{taylor}
\end{eqnarray}
Note that in our case ${ {d^2u} \over {dy^2}}= -u$ and, moreover,  
$\delta y$ performs an Ornestein-Ulenbeck process, i.e. 
a Gaussian process with 
$\langle\delta y\rangle=0$, 
 $\langle(\delta y)^2\rangle=D_0/K$ and a characteristic relaxation 
time $\tau=1/(2K)$.
Therefore, from Eq.~(\ref{taylor}) one sees that the center of mass 
feels a renormalized velocity field 
$u(Y) \to u(Y)(1-\langle(\delta y)^2\rangle/8)$
and a renormalized fluctuation part, i.e. with a larger $D_0$.
Now, because of the Taylor formula, $D=D_0/2+U^2/[2 (D_0/2)]$,
(note that $D_0\to D_0/2$ in the particle limit defined by Eq.~(\ref{center}) 
it is easy
to see that $D$  has negative correction for finite values of $K$.\\
For a generic time independent shear flow we expect the same 
qualitative behavior. This fact can be grasped noting that
for  $u(y)=\sum_k u_k exp(ik y)+ c.c.$, $u_k$ being the Fourier transform of 
 $u(y)$ and $c.c.$ stands for the complex conjugate,  
the center of mass experiences a normalized velocity field
$\sum_k u_k(1-k^2\langle(\delta y)^2/8\rangle)$. Exploiting the Zeldovich 
formula \cite{Z82}, $D\sim \sum_k |u_k|^2/(D_0 k^2)$, in all the realistic 
cases having rapidly vanishing values of $|u_k|^2$ in the limit $k\to \infty$,
the above argument reported for the Kolmogorov flow still holds.

It is worth observing that, for  time dependent shear flows, 
the increasing fluctuating part could give rise to transport
enhancement via the interference mechanism identified in Ref.~\cite{MV97}.
Indeed, in the presence of anticorrelated regions of the velocity field
(i.e., where the velocity autocorrelation function is negative),
the enhanced molecular
diffusivity can be advantageous to escape from the anticorrelated
regions which slow down the diffusing particle. An enhancement
in the diffusion coefficient might occur in this case.

In conclusion, two different mechanisms leading to transport
reduction have been identified for two paradigmatic flows
with closed and open streamlines, respectively.
 For stationary flows with closed streamlines transport reduction
is due to the emergence of dynamically generated barriers.
 For stationary flows with open streamlines the mechanism
giving rise to the observed reduction of transport is triggered by a
renormalized (enhanced) molecular diffusivity. 
In virtue of such enhanced molecular 
diffusivity, the bead center of mass forget its past evolution faster than 
in the case of particles, a fact that reduces the eddy diffusivity.\\
It is an open question whether, in the presence of
time-dependent shear flows, the same renormalization of the bare molecular
diffusivity might give rise to transport enhancement via
the interference mechanism identified in Ref.~\cite{MV97}.

\vskip 0.2cm
\noindent {\bf Acknowledgements}
\\
Helpful discussions and suggestions by
R.~Festa are acknowledged. 
This work has been partially supported by the INFM project GEPAIGG01 (A.M.).
M.D.L. and  A.V. acknowledge support from the INFM {\em Center for
Statistical Mechanics and Complexity} (SMC).
Simulations were performed at CINECA (INFM Parallel Computing Initiative).

\end{document}